\pdfoutput=1

\documentclass[11pt,twoside,a4paper,cmspaper,final,collab]{cms-tdr}
\def\svnVersion{488899P}\def\cmsCernNoTag{CERN-EP-2018-056}\def\cmsCernDate{\today}\def\cmsMessage{Published in Physics Letters B as \href{http://dx.doi.org/10.1016/j.physletb.2019.01.058}{\doi{10.1016/j.physletb.2019.01.058.}}}

\begin{document}\cmsNoteHeader{HIN-16-015}

\hyphenation{had-ron-i-za-tion}
\hyphenation{cal-or-i-me-ter}
\hyphenation{de-vices}
\RCS$HeadURL: svn+ssh://svn.cern.ch/reps/tdr2/papers/HIN-16-015/trunk/HIN-16-015.tex $
\RCS$Id: HIN-16-015.tex 485607 2019-01-02 21:05:52Z alverson $

\ifthenelse{\boolean{cms@external}}{\providecommand{\cmsLeft}{top\xspace}}{\providecommand{\cmsLeft}{left\xspace}}
\ifthenelse{\boolean{cms@external}}{\providecommand{\cmsRight}{bottom\xspace}}{\providecommand{\cmsRight}{right\xspace}}

\newcommand{\mumu}{\ensuremath{\PGmp\PGmm}\xspace}
\newcommand{\B}{\ensuremath{\cmsSymbolFace{B}}\xspace}
\newcommand{\rpa}{\ensuremath{R_{\pPb}}\xspace}
\newcommand{\ycm}{\ensuremath{y_{\text{CM}}}\xspace}
\newcommand{\pp}{{\ensuremath{\Pp\Pp}}\xspace}
\newcommand{\pPb}{\ensuremath{\Pp\text{Pb}}\xspace}
\newcommand{\PbPb}{\ensuremath{\text{PbPb}}\xspace}
\newcommand{\sqrts}{\ensuremath{\sqrt{s}}\xspace}
\newcommand{\Lxy}{\ensuremath{L_{xy}}\xspace}
\newcommand{\ctau}{\ensuremath{\ell_{\PJGy}}\xspace}
\newcommand{\AuAu}{\ensuremath{\text{AuAu}}\xspace}
\newcommand{\pA}{\ensuremath{\Pp\text{A}}\xspace}
\newcommand{\PgUn}{\ensuremath{\Upsilon\text{(nS)}}\xspace}

\cmsNoteHeader{HIN-16-015}
\title{Measurement of prompt \Pgy\ production cross sections in proton-lead and proton-proton collisions at  \texorpdfstring{$\sqrtsNN=5.02\TeV$}{sqrt(s[NN]) = 5.02 TeV}}

\date{\today}

\abstract{
Measurements of prompt \Pgy\ meson production cross sections in proton-lead (\pPb)
and proton-proton (\pp) collisions at a nucleon-nucleon center-of-mass energy of {\sqrtsNN = 5.02\TeV} are reported.
The results are based on \pPb and \pp data collected by the CMS experiment at the LHC,
corresponding to integrated luminosities of {34.6\nbinv} and {28.0\pbinv},
respectively. The nuclear modification factor \rpa is measured for prompt \Pgy\ in the
transverse momentum range $4 < \pt < 30$\GeVc and the center-of-mass rapidity range
$-2.4 < \ycm < 1.93$. The results on \Pgy\ \rpa are compared to the corresponding modification factor
for prompt \PJGy\ mesons. The results point to different nuclear effects at play in the production of the excited charmonium state compared to the ground state, in the region of backward rapidity and for $\pt<10\GeVc$.
}

\hypersetup{%
pdfauthor={CMS Collaboration},%
pdftitle={Measurement of prompt psi(2S) production cross sections in proton-lead and proton-proton collisions at sqrt(s[NN]) = 5.02 TeV},%
pdfsubject={CMS},%
pdfkeywords={CMS, pPb, psi(2S)}%
}

\maketitle

\section{Introduction}
The study of quarkonium production in nuclear collisions has a long and rich history,
dating back to the original proposal by Matsui and Satz
predicting \PJGy\ suppression in heavy ion collisions due to Debye screening in the
quark-gluon plasma (QGP)~\cite{Matsui:1986dk}. After this proposal, the first \PJGy\ measurements
in heavy ion collisions were performed in fixed target experiments at the CERN SPS, at a nucleon-nucleon
center-of-mass energy of {$\sqrtsNN \approx 20$\GeV}~\cite{Baglin:1990iv,Abreu:1997jh}.
A similar level of suppression was later observed in  gold-gold collisions
at the BNL RHIC, at {$\sqrtsNN=200\GeV$}~\cite{Adare:2006ns,Adare:2011yf}.
At the CERN LHC, the production of charmonium (\PJGy, \Pgy) and bottomonium (\PgUa, \PgUb, \PgUc)
states has been studied in lead-lead (PbPb) collisions at {$\sqrtsNN=2.76$} and
{5.02\TeV}~\cite{Abelev:2013ila,Khachatryan:2016ypw,Khachatryan:2014bva,Chatrchyan:2012lxa,Khachatryan:2016xxp,Sirunyan:2016znt,Sirunyan:2017lzi},
bringing new elements for the understanding of the medium produced in high-energy heavy ion collisions.

The quarkonium yields can be modified in heavy ion collisions because of several effects~\cite{Kumar:2014kfa}, including suppression~\cite{Matsui:1986dk} and recombination of charm quark pairs~\cite{Braun-Munzinger:2015hba} inside the hot matter (QGP), and cold nuclear matter effects. An unambiguous interpretation of the nucleus-nucleus LHC results requires a quantitative understanding of cold nuclear matter effects. The nuclear parton distribution functions (nPDFs) are known to be different from those in a free proton~\cite{Ferreiro:2013pua,Vogt:2015uba}. In addition, gluon radiation induced by multiple parton scattering in the nucleus leads to transverse momentum (\pt) broadening and coherent energy loss, resulting in a significant quarkonium suppression in nuclear collisions at all available energies~\cite{Arleo:2012hn,Arleo:2013zua}. These phenomena are best studied in proton-nucleus collisions, in which hot medium effects, such as those due to the QGP, are likely to be limited.

Many charmonium production measurements have been performed in proton-induced collisions, on several nuclei, at the SPS~\cite{Badier:1983dg,Abreu:1998ee,Alessandro:2003pc,Arnaldi:2010ky}, HERA~\cite{Abt:2008ya}, and Tevatron~\cite{Leitch:1999ea}. A global analysis of the fixed-target measurements can be found in Ref.~\cite{Lourenco:2008sk}. At RHIC, \PJGy\ and \Pgy\ data in deuteron-gold collisions have been reported by the PHENIX~\cite{Adler:2005ph} and STAR~\cite{Adamczyk:2013poh} Collaborations. At the LHC, the cross section of \PJGy\ in proton-lead (\pPb) collisions at {$\sqrtsNN=5.02\TeV$}
has been measured by the ALICE~\cite{Abelev:2013yxa,Adam:2015iga}, ATLAS~\cite{Aad:2015ddl}, CMS~\cite{Sirunyan:2017mzd}, and LHCb~\cite{Aaij:2013zxa} Collaborations. A significant suppression of the prompt \PJGy\ yield in \pPb collisions has been observed at forward rapidity ($y$) and low \pt, while no strong nuclear effects are reported at backward rapidity, where forward and backward indicate, respectively, the directions of the proton and Pb beams. Measurements of the \PgUa\ in \pPb collisions at {$\sqrtsNN=5.02\TeV$} have also been performed by the ALICE~\cite{Abelev:2014oea}, ATLAS~\cite{Aaboud:2017cif}, and LHCb~\cite{Aaij:2014mza} experiments, indicating that the \PgUa\ state is less suppressed than the \PJGy.

Additional information can be obtained by studying the behavior of the excited states, which are less tightly bound compared to the ground states and might suffer stronger suppression in heavy ion collisions. At the LHC, the ALICE~\cite{Adam:2015isa} and CMS~\cite{Sirunyan:2016znt} Collaborations have reported a stronger suppression of the \Pgy\ state compared to the \PJGy\ in \PbPb collisions. In \pPb collisions, ALICE~\cite{Abelev:2014zpa}, ATLAS~\cite{Aaboud:2017cif}, and LHCb~\cite{Aaij:2016eyl} data show that \Pgy\ suppression, integrated over transverse momentum, is more pronounced than that of the \PJGy. This observation suggests final-state effects for the excited states, possibly due to inelastic interactions with the medium produced in \pPb collisions~\cite{Ferreiro:2014bia}. In the bottomonium sector, the double yield ratios  \PgUc/\PgUa\ and \PgUb/\PgUa\ in \pPb relative to \pp collisions have been measured by CMS~\cite{Chatrchyan:2013nza} and found to be less than unity, again indicating stronger final-state effects in the production of excited quarkonium states.

This Letter reports measurements of the cross sections for prompt \Pgy\ production in \pp and \pPb collisions at {$\sqrtsNN=5.02\TeV$}, with the \Pgy\ decaying to $\mu^+\mu^-$, over the \pt range 4--30\GeVc and center-of-mass rapidity range $-2.4<\ycm<1.93$. The data were collected with the CMS detector at the LHC, in 2013 for the \pPb sample and in 2015 for the \pp sample, corresponding to integrated luminosities of {$34.6\pm 1.2\nbinv$}~\cite{CMS-PAS-LUM-13-002}  and {$28.0\pm 0.6\pbinv$}~\cite{CMS-PAS-LUM-16-001}, respectively. In addition, the modification of quarkonium production in \pPb collisions is quantified by the nuclear modification factor, \rpa, which is defined as the ratio of the cross sections in \pPb and \pp collisions divided by the Pb mass number, $A=208$. The \Pgy\ \rpa is determined as a function of \ycm and \pt, and is compared to that of the \PJGy\ mesons measured at the same center-of-mass energy~\cite{Sirunyan:2017mzd}. This is the first measurement of \Pgy\ \rpa at {$\sqrtsNN=5.02 \TeV$} in differential bins of \pt and \ycm that uses the \pp data measured at the same center-of-mass energy.

\section{The CMS detector}
The central feature of the CMS apparatus is a superconducting solenoid of 6\unit{m} internal diameter, providing a magnetic field of 3.8\unit{T}. Within the solenoid volume there are a silicon pixel and strip tracker, a lead tungstate crystal electromagnetic calorimeter, and a brass and scintillator hadron calorimeter, each composed of a barrel and two endcap sections. Forward calorimeters extend the pseudorapidity ($\eta$) coverage provided by the barrel and endcap detectors. The forward hadron (HF) calorimeter uses steel as absorber and quartz fibers as the sensitive material. The two halves of the HF are located at $\pm$11.2\unit{m} from the nominal interaction point. Together they provide coverage in the range $3.0 < \abs{\eta} < 5.2$ and also serve as luminosity monitors. Muons are detected in gas-ionization chambers embedded in the steel flux-return yoke outside the solenoid, in the range $\abs{\eta} < 2.4$, with detection planes made using three technologies: drift tubes, cathode strip chambers, and resistive-plate chambers. Matching muons to tracks measured in the silicon tracker results in a relative \pt resolution for typical muons in this analysis of 1.3--2.0\% in the barrel and better than 6\% in the endcaps~\cite{Chatrchyan:2012xi}. A more detailed description of the CMS detector, together with a definition of the coordinate system used and the relevant kinematic variables, can be found in Ref.~\cite{Chatrchyan:2008zzk}.

\section{Event selection}
The \pPb collisions at {$\sqrtsNN=5.02$\TeV} correspond to a proton beam energy of 4\TeV and a lead beam energy of 1.58\TeV per nucleon. The proton beam traveled in the $-\eta$ direction (in the detector coordinate system) in the first part of the run, corresponding to an integrated luminosity {$\mathcal{L}_{\text{int}}=20.7$\nbinv}, and in the opposite direction in the second part of the run, corresponding to {$\mathcal{L}_{\text{int}}=13.9$\nbinv}. Particles emitted at $\abs{\eta_{\text CM}}=0$ in the nucleon-nucleon center-of-mass frame are detected in the laboratory frame at $\eta_{\text{lab}} = \pm 0.465$ depending on the proton beam direction. In this paper, data for half of the run were reflected so that positive $\eta$ always corresponds to the direction of the proton beam. The \pp data set, collected at the same collision energy as the \pPb sample, corresponds to $\mathcal{L}_{\text{int}} = 28.0$\pbinv. In the \pp sample, the dimuons from \Pgy\ decays are reconstructed within $\abs{\ycm} < 2.4$ and in the same \pt range as the \pPb data, $4 < \pt < 30$\GeVc.

Inelastic hadronic collisions are selected by requiring  at least one HF tower with more than 3\GeV of total energy in each of the two HF calorimeters. This is not required in \pp collisions, which suffer less from photon-induced interactions compared to \pPb collisions. The \pp and \pPb events are further selected to have at least one reconstructed primary vertex composed of two or more associated tracks, excluding the two muons, within 25\cm from the nominal interaction point along the beam axis and within 2\cm in the transverse plane. To reject the beam-gas background events, the fraction of good-quality tracks associated with the primary vertex is required to be larger than 25\% when there are more than 10 tracks per event.

The dimuon events are selected by the level-1 trigger, a hardware-based trigger system requiring two muon candidates in the muon detectors with no explicit limitations on their \pt or $\eta$. In the offline analysis, muons are required to be within the following kinematic regions, which ensure single-muon reconstruction efficiencies above 10\%:
\begin{linenomath}
\begin{equation}
\label{eqn:acccut}
\begin{aligned}
   &\pt^{\mu}>3.3\GeVc && \text{ for $\abs{\eta_{\text{lab}}^{\mu}}<1.2$},\\
   &\pt^{\mu}>(4-[1.1\,\abs{\eta_{\text{lab}}^{\mu}}])\GeVc &&\text{ for $1.2<\abs{\eta_{\text{lab}}^{\mu}}<2.1$},\\
   &\pt^{\mu}>1.3\GeVc &&\text{ for $2.1<\abs{\eta_{\text{lab}}^{\mu}}<2.4$}.
\end{aligned}
\end{equation}
\end{linenomath}
The oppositely charged muon pairs are further selected to originate from a common
vertex with a $\chi^2$ probability greater than 1\%, and to survive standard identification
criteria~\cite{Chatrchyan:2012xi}. In order to remove cosmic-ray muons, the transverse and longitudinal
distances of closest approach between the muon trajectory and the reconstructed primary vertex are required
to be less than 0.3\cm and 20\cm, respectively. All of these selections are common to both the \pp and the \pPb data sets.

\section{Yield extraction}
\label{sec:YieldExtraction}
The prompt \Pgy\ signal extraction procedure is similar to that in previous CMS analyses~\cite{Khachatryan:2014bva,Sirunyan:2016znt}, and is the same for the \pp and \pPb data sets. The dimuon mass distribution is fitted with signal (including both the \PJGy\ and \Pgy\ resonances) and background contributions using an extended unbinned maximum likelihood procedure~\cite{roofit}. While the scope of the present Letter is the measurement of the prompt \Pgy\ production, the \PJGy\ resonance was included in the \Pgy\ fit to increase the peak fitting stability. It has been checked that the prompt \PJGy\ results obtained in this analysis agree with the ones in Ref.~\cite{Sirunyan:2017mzd}, which used a different method for separating prompt from nonprompt charmonia. The prompt \PJGy\ results of Ref.~\cite{Sirunyan:2017mzd} are the ones used in this Letter for comparison with the new analysis of \Pgy\ production. Both resonance shapes are modeled by the weighted sum of a Crystal Ball (CB)~\cite{bib:CrystalBall} and a Gaussian function. The CB function, $g_\text{CB}(m)$, combines a Gaussian core and a power law tail with exponent $n$, accounting for energy loss due to final-state photon radiation, with a parameter $\alpha$ defining the transition between the Gaussian and
 the power law functions:
 \ifthenelse{\boolean{cms@external}}{
   \begin{equation}
    \label{eq:cb}
     \begin{split}
       &g_{\text{CB}}(m) = \\
       &\left\{
      \begin{aligned}
        &\frac{N}{\sqrt{2\pi}\,\sigma_{\text{CB}}}\exp\biggl(-\frac{(m-m_0)^2}{2\,\sigma_{\text{CB}}^2}\biggr),& \text{for $\frac{m-m_0}{\sigma_{\text{CB}}}>-\alpha$};\\
        &\frac{N}{\sqrt{2\pi}\,\sigma_{\text{CB}}}\biggl(\!\frac{n}{\lvert\alpha\rvert}\!\biggr)^n\exp\biggl(-\frac{\lvert\alpha\rvert^2}{2}\biggr)\\
        &\quad\times\biggl(\frac{n}{\lvert\alpha\rvert}-\lvert\alpha\lvert-\frac{m-m_0}{\sigma_{\text{CB}}}\biggr)^{-n},& \text{for $\frac{m-m_0}{\sigma_{\text{CB}}}<-\alpha$}.&
     \end{aligned}
  \right.
  \end{split}
   \end{equation}
 }{
\begin{linenomath}
   \begin{equation}
    \label{eq:cb}
    g_{\text{CB}}(m) = \left\{
      \begin{aligned}
        &\frac{N}{\sqrt{2\pi}\,\sigma_{\text{CB}}}\exp\biggl(-\frac{(m-m_0)^2}{2\,\sigma_{\text{CB}}^2}\biggr),& \text{for $\frac{m-m_0}{\sigma_{\text{CB}}}>-\alpha$};\\
        &\frac{N}{\sqrt{2\pi}\,\sigma_{\text{CB}}}\biggl(\!\frac{n}{\lvert\alpha\rvert}\!\biggr)^n\exp\biggl(-\frac{\lvert\alpha\rvert^2}{2}\biggr)\biggl(\frac{n}{\lvert\alpha\rvert}-\lvert\alpha\lvert-\frac{m-m_0}{\sigma_{\text{CB}}}\biggr)^{-n},& \text{for $\frac{m-m_0}{\sigma_{\text{CB}}}<-\alpha$}.&
  \end{aligned}\right.
   \end{equation}
\end{linenomath}
}
The CB and Gaussian functions have independent widths, $\sigma_\text{CB}$ and $\sigma_\text{G}$,
to accommodate the dependence of the dimuon invariant mass resolution on the dimuon rapidity and \pt,
but share a common mean $m_0$ representing the \PJGy\ mass. The signal models used for the \Pgy\ and \PJGy\ share the same $\alpha$ and $n$ parameters. The mean and the width of the Gaussian model for the \Pgy\ are obtained from those of the \PJGy, scaled by their mass ratio ($m_{\Pgy}/m_{\PJGy} = 1.1903$~\cite{pdg:2016}). The following parameters are left free in the fit: $m_{0}$, $\sigma_{\text{CB}}$, $\sigma_{\text{G}}$, $N_{\PJGy}$ (the \PJGy\ yield), $N_{\Pgy}$ (the \Pgy\ yield), $f$ (the relative contribution of the Gaussian and CB functions), and $\alpha$. Based on simulation studies, the value of the parameter $n$ is  fixed to $2.1$. The parameter $\alpha$ is left free, covering the variation of signal shapes in each of the \pt and rapidity bins. The same value for $f$ is used in the definition of the \Pgy\ and \PJGy\ signal shapes. The underlying background is described by a Chebyshev polynomial of degree $N~(1 \leq N \leq 3)$. The degree of the Chebyshev polynomial is obtained in each $(\pt, \ycm)$ bin of the analysis using a log-likelihood ratio test. Several alternative fitting procedures have been tested, and the yield variations with respect to the nominal result are included as a systematic uncertainty, as explained in Section~\ref{sec:Syst}. The two \pPb data sets, corresponding to each proton beam direction, are merged and analyzed together after having verified that the independent results are compatible with each other.

The \PJGy\ and \Pgy\ coming from \cPqb{} hadron decays (nonprompt charmonia) are evaluated
using the displacement of the \mumu vertex from the primary collision vertex. This secondary \mumu vertex is characterized by the pseudo-proper decay length,
\begin{linenomath}
   \begin{equation}
      \ctau = \Lxy \, \frac{m_{\PJGy}} {\pt},
   \end{equation}
\end{linenomath}
where $\Lxy$ is the transverse distance between the primary and dimuon vertices, $m_{\PJGy}$ is the mass of the \PJGy~\cite{pdg:2016} (assumed for all muon pairs), and \pt is the transverse momentum of the dimuon.
Prompt charmonia are selected by requiring \ctau to be smaller than a threshold
value $l_0$, which is tuned using simulation separately for the \pp and \pPb collisions systems, in order to keep 90\% of the total prompt charmonia~\cite{Sirunyan:2016znt}. Since the \ctau resolution improves with increasing dimuon \pt, the $l_0$ values also depend on \pt. The $l_0$ values used remove between 70\% and 90\% (80\% and 90\%) of the nonprompt \Pgy\ in \pPb (\pp) collisions, from low- to high-\pt.

The yields so obtained have a small nonprompt contamination, which is corrected
using the number of simulated events  passing and failing the \ctau
threshold criteria~\cite{Sirunyan:2016znt}. The full value of this correction
is also propagated as a source of systematic uncertainty.

\begin{figure}[ht!]
\centering
\includegraphics[width=0.49\textwidth]{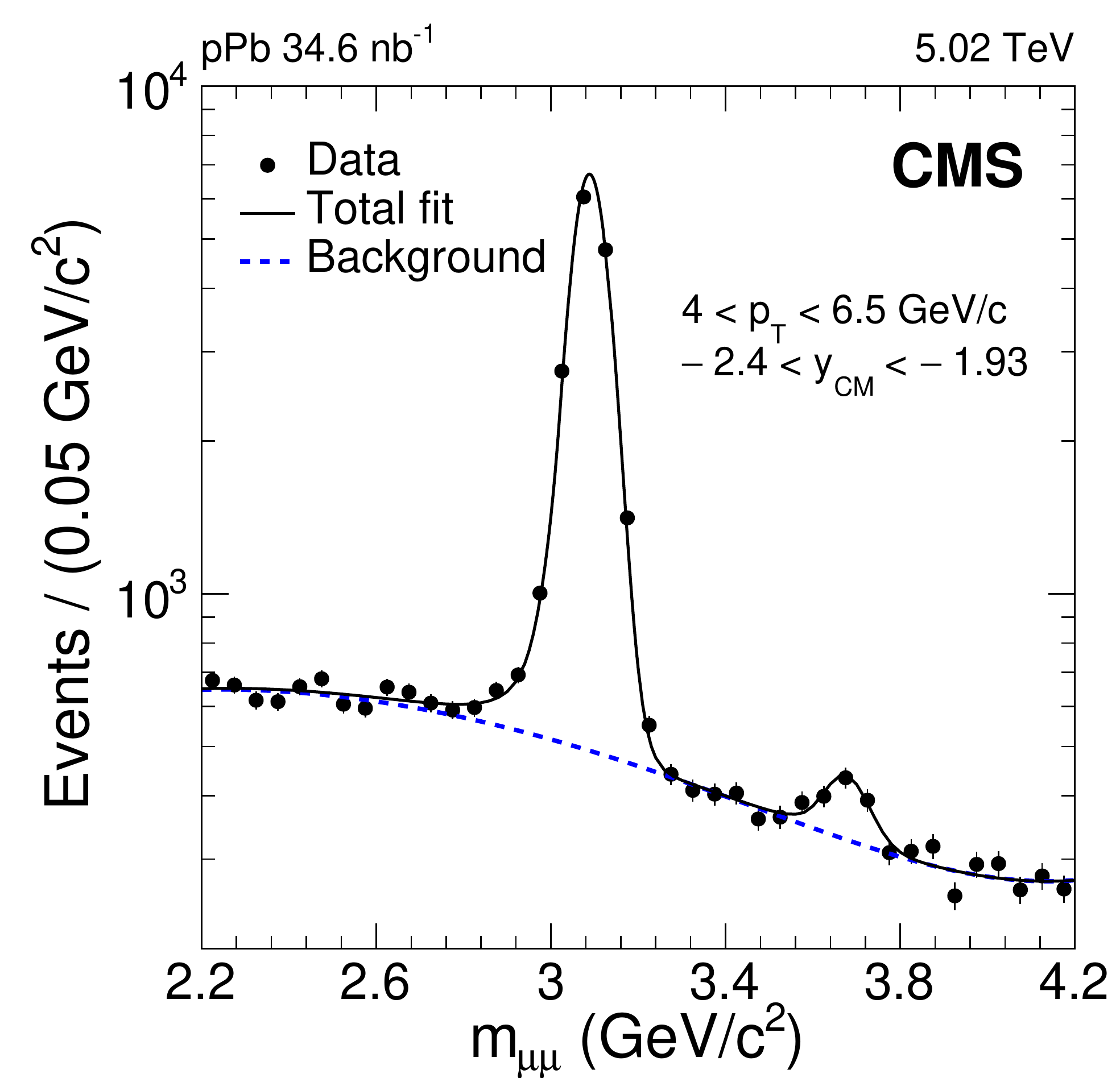}
\includegraphics[width=0.49\textwidth]{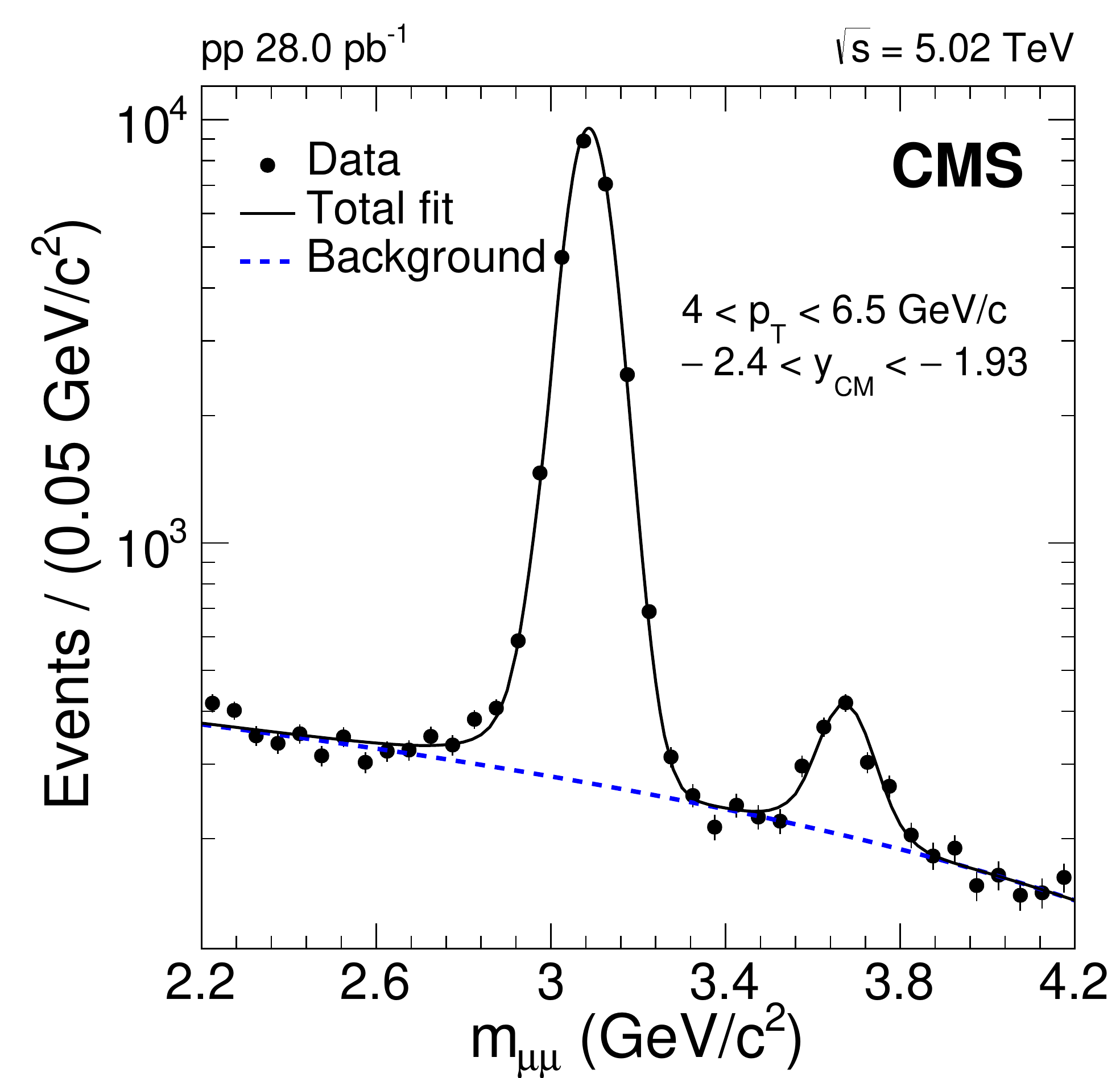}
\caption{\label{Fig:YieldPPb} Dimuon invariant mass distribution
showing the \PJGy\ and \Pgy\ peaks in \pPb (\cmsLeft panel) and \pp (\cmsRight panel) data,
after applying the \ctau selection, for {$-2.4 < \ycm < -1.93$} and {$4 < \pt < 6.5$\GeVc}.
   The fits to the distributions are also shown.
   }
\end{figure}

Figure~\ref{Fig:YieldPPb} shows the dimuon invariant mass distributions with the \PJGy\ and \Pgy\ peaks in \pPb and \pp data, for one of the bins with the smallest signal-to-background ratio (lowest \pt and most backward $y$), after applying the \ctau selection, together with the curves resulting from the fits. The raw prompt \Pgy\ yield extracted in these bins is $\sim$300 for the \pPb, and twice as much for the \pp samples.

\section{Acceptance and efficiency corrections}
\label{sec:AccAndEff}
Simulated events are used to obtain the acceptance and efficiency correction factors
for the measured \Pgy\ yields. The events are generated using
\PYTHIA 6.424~\cite{Sjostrand:2006za} for \pPb collisions and \PYTHIA 8.209~\cite{Sjostrand:2014zea}
for \pp collisions.
The generated particles in the \pPb Monte Carlo (MC) are boosted by the $\beta$ of the center-of-mass system
in the laboratory frame, resulting in $\Delta y = \pm 0.465$.
The prompt \PJGy\ and \Pgy\ are assumed
to be produced unpolarized in both \pp and \pPb collisions, which is supported by measurements
in \pp collisions at {$\sqrts=7$\TeV}~\cite{Chatrchyan:2013cla,Khachatryan:2016vxr}.
The final-state QED radiation of the decay muons is
simulated using \PHOTOS 215.5~\cite{Barberio:1993qi}.
Finally, the CMS detector response is simulated using \GEANTfour~\cite{Agostinelli:2002hh}.

The acceptance in a given $(\pt, \ycm)$ bin is defined as the fraction of generated \Pgy\ mesons
resulting in a detectable muon pair, \ie, passing the single muon selection criteria defined in Eq.~(\ref{eqn:acccut}).
The  efficiency is given by the fraction of generated and detectable muon pairs that result in a reconstructed muon pair, also passing the trigger and offline selections. The combined \Pgy\ acceptance and efficiency for \pPb (\pp) results ranges between 13\% (14\%) in the lowest \pt region ($4.0 < \pt < 6.5$\GeVc) and 66\% (67\%) in the highest \pt region ($10.0 < \pt < 30.0$\GeVc). It is maximum in the overlap region between barrel and endcap detectors, and decreases to minimum values for $|\ycm|>2$.

The individual components of the simulation-based dimuon efficiency (track reconstruction, muon identification and selection, and triggering) are probed using single muons from \PJGy\ meson decays in both simulated and collision data, with the \textit{tag-and-probe} (T\&P) method~\cite{Khachatryan:2010xn,Sirunyan:2017mzd}. The data-to-simulation ratios of single muon efficiencies obtained from T\&P, calculated as a function of muon $\eta$ and \pt, are used as weights (\textit{scale factors}) to correct event-by-event the dimuon efficiency. The effect of this correction on the dimuon efficiency is similar for \pp and \pPb events and ranges from 0.99 to 1.33, with larger corrections at lower \pt and more forward/backward rapidities.

\section{Systematic uncertainties}
\label{sec:Syst}
The following sources of point-to-point systematic uncertainties are considered for both \pp and \pPb data:
the yield extraction method, given the choice of signal and background models;
acceptance and efficiency corrections, including T\&P scale factors and
the possible difference in the dimuon \pt spectrum between data
and simulation; and the nonprompt charmonia contamination.
The evaluation of these systematic uncertainties is described below.

\begin{itemize}
\item \textbf{Signal shape variation.} This systematic uncertainty
is obtained by changing the fitting constraints on the CB shape parameters.
In the nominal fits, the CB parameter $n$ is fixed from simulation, $n=2.1$,
and the parameter $\alpha$ is left free. In a first variation, the parameter $\alpha$ is fixed
to the MC-based value 1.7, and $n$ is left free.
A second variation consists in fixing
both parameters to the values extracted from MC, $n=2.1$ and $\alpha=1.7$.
The maximum  difference, in each analysis bin, between the yields extracted with the nominal fit and either of these variations,
is taken as a systematic uncertainty.
\item \textbf{Background shape variation.}
  The degree $N$ of the Chebyshev polynomial
  is changed to $N+1$ if $N<3$, and to $N-1$ if $N=3$. The systematic uncertainty is
      estimated as the absolute difference in the
  yields with respect to the nominal case.
\item \textbf{Simulated dimuon \pt spectrum.}
  The acceptance and efficiency correction factors depend on
    the shape of the simulated
    \Pgy\ \pt distribution; the difference
  between data and simulation is a source of systematic uncertainty. The ratio of the corrected yields
  in data and simulation is evaluated as a function of \pt, for each rapidity bin.
  Continuous weighting factors are obtained by fitting these data-to-simulation ratios with a linear
  function in \pt.
  The acceptance and efficiency values are evaluated again after weighting the \pt
  distribution of the generated \Pgy\ mesons in the simulation by the function obtained.
  The difference between the reweighed simulation and the nominal yields is taken as a systematic uncertainty.
\item \textbf{T\&P scale factors.}
  The statistical uncertainty on the T\&P scale factors, as well as systematic uncertainties in their derivation,
      are accounted for as a systematic uncertainty. These uncertainties are further described
  in Ref.~\cite{Sirunyan:2017mzd}.
\item \textbf{Nonprompt contamination. }
  The difference between the prompt meson yields with and without the correction for the residual nonprompt
contamination is propagated as a systematic uncertainty.
\end{itemize}

The systematic uncertainty in the yield extraction, determined by summing the uncertainties from the signal and background shape variations in quadrature, is in the range 1.0--22\% and 0.4--6.5\% for the \pPb and \pp results, respectively. This large range is mostly driven by the variation of the signal over background ratio across the analysis bins. The dominant part in this uncertainty is from background modeling, with an average over all the \pt and \ycm bins of 3.9 (2.4)\% and a maximum of 22 (4.9)\% for the \pPb (\pp) results. The average signal modeling contribution is 1.9\% and 1.4\%, for the \pPb and \pp results, respectively.

The systematic uncertainty in the acceptance and efficiency correction factors, combining the dimuon \pt spectrum reweighting and T\&P uncertainties, lies within 2.0--7.4\% and 1.9--8.1\% for the \pPb and \pp results, respectively. The prompt selection method induces an uncertainty of 4.5--15\% (0.2--8.0\%) for the \pPb (\pp) results, with the maximum value in the lowest \pt bin in the backward $y$ region. The total systematic uncertainty, calculated as the sum in quadrature of the individual uncertainties, is in the range 5--26\% (4--11\%) for the \pPb (\pp) cross section measurements. In general, the uncertainty is minimum in the highest \pt bins in the central $y$ region, and increases when going to lower \pt and forward/backward $y$ regions. For the \rpa measurement, the \pp and \pPb uncertainties are considered independent and added in quadrature point-by-point. The total systematic uncertainty in this case is in the range 5.0--22.6\%.

The global (common to all points) systematic uncertainty due to the uncertainty in the integrated luminosity of the pPb and \pp data sets is 3.5\%~\cite{CMS-PAS-LUM-13-002} and 2.3\%~\cite{CMS-PAS-LUM-16-001}, respectively.  For the cross section results, this uncertainty is not included to the point-to-point uncertainties, while for the \rpa results the two numbers are added in quadrature and displayed as a box around $\rpa=1$. The dominant source of uncertainty for the \rpa measurement is the statistical uncertainty in the \pPb data.

\section{Results}
\label{Sec:Results}

The prompt \Pgy\ production cross section (multiplied by the \Pgy\ branching fraction to $\mumu$, $\mathcal{B}(\Pgy\to\mumu)$) in the dimuon decay
channel is determined as
\begin{linenomath}
   \begin{equation}
\mathcal{B}(\Pgy\to\mumu)\frac{\rd^2\sigma}{\rd\pt\,\rd\ycm}
= \frac{N^{\Pgy}_{\text{fit}}/(\text{acc}\,\varepsilon)}{\mathcal{L}_{\text{int}}\,\Delta\pt\,\Delta\ycm},
\label{eqn:crosssection}
   \end{equation}
\end{linenomath}
where $N^{\Pgy}_{\text{fit}}$ is the extracted raw yield of prompt
\Pgy\ mesons in a given $(\pt,\ycm)$ bin, $(\text{acc}\,\varepsilon)$ is
the product of the dimuon acceptance and efficiency described in Section~\ref{sec:AccAndEff},
and $\Delta\pt$ and $\Delta\ycm$ are the widths of the kinematic bin considered.

\begin{figure}[ht!]
  \centering
  \includegraphics[width=0.48\textwidth]{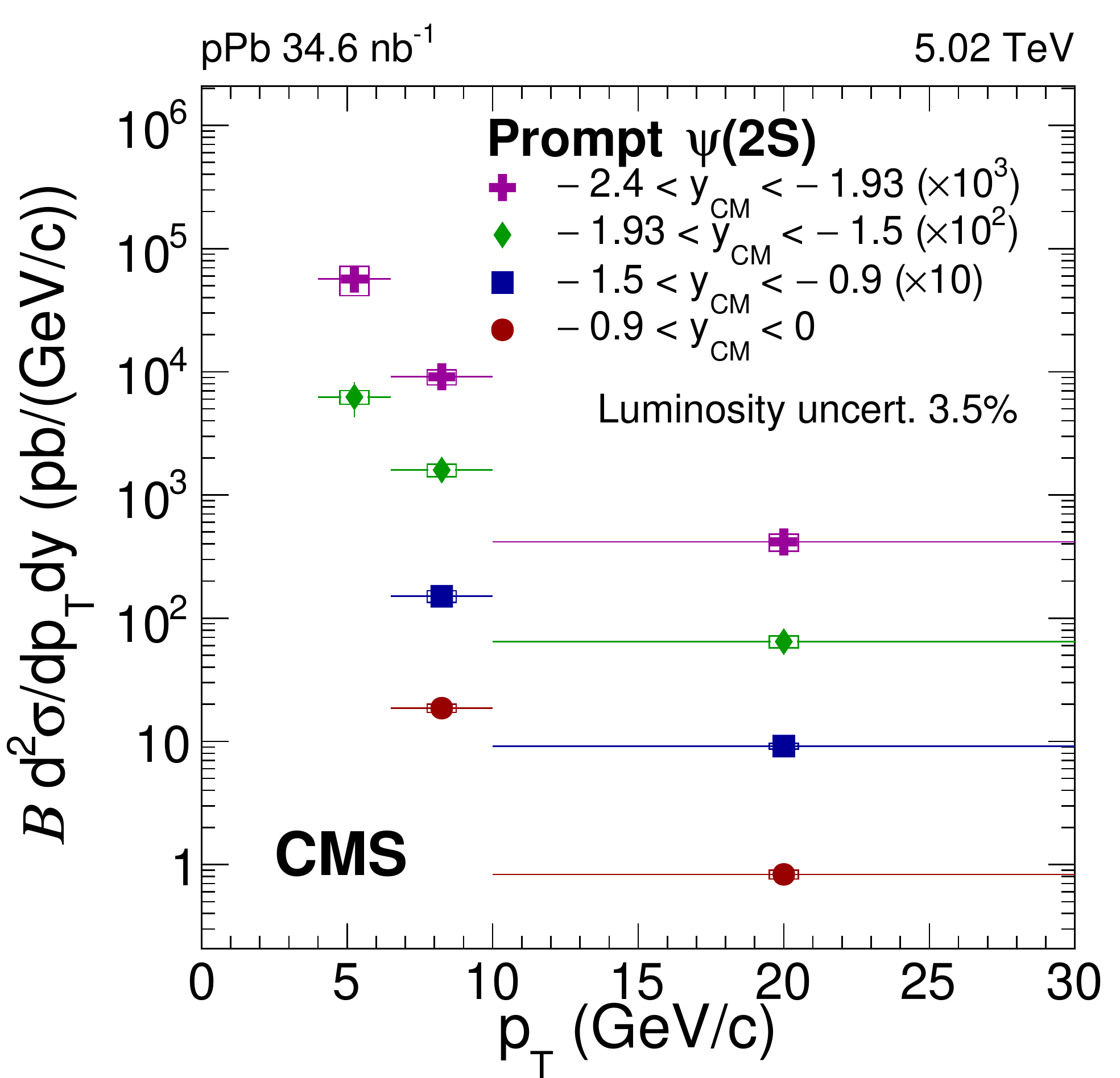}
  \includegraphics[width=0.48\textwidth]{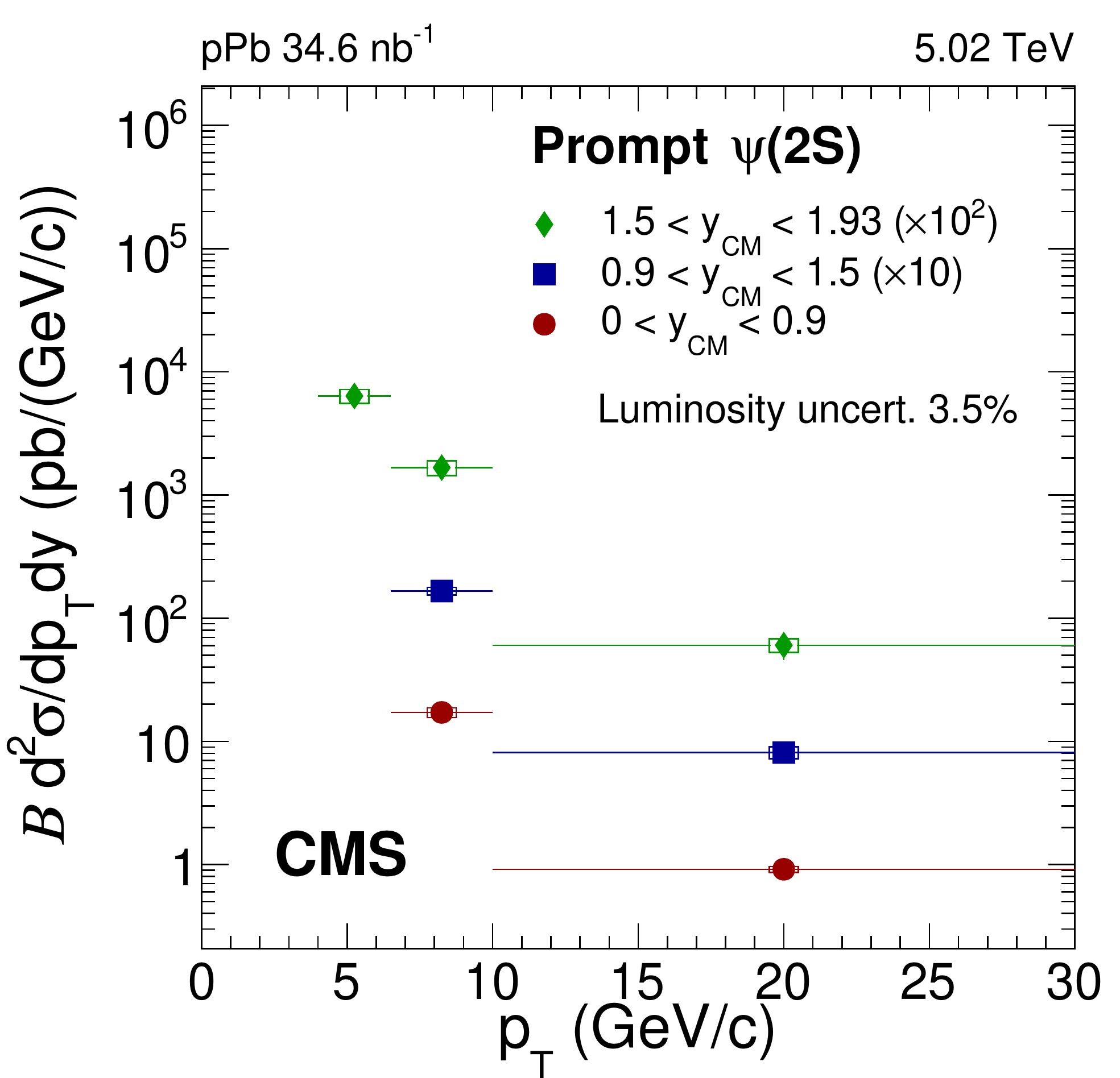}
  \caption{\label{Fig:sigma_ppb} Differential cross section (multiplied by the dimuon branching fraction)
    of prompt \Pgy\ production in \pPb collisions at {$\sqrtsNN=5.02\TeV$}, as a function of \pt, in several rapidity bins and
    separately for backward (\cmsLeft panel) and forward (\cmsRight panel) rapidity regions.
    Statistical and systematic uncertainties are represented with error bars and boxes, respectively. The fully correlated luminosity
    uncertainty of 3.5\% is not included in the point-by-point uncertainties.}
\end{figure}

\begin{figure}[ht]
  \centering
  \includegraphics[width=0.48\textwidth]{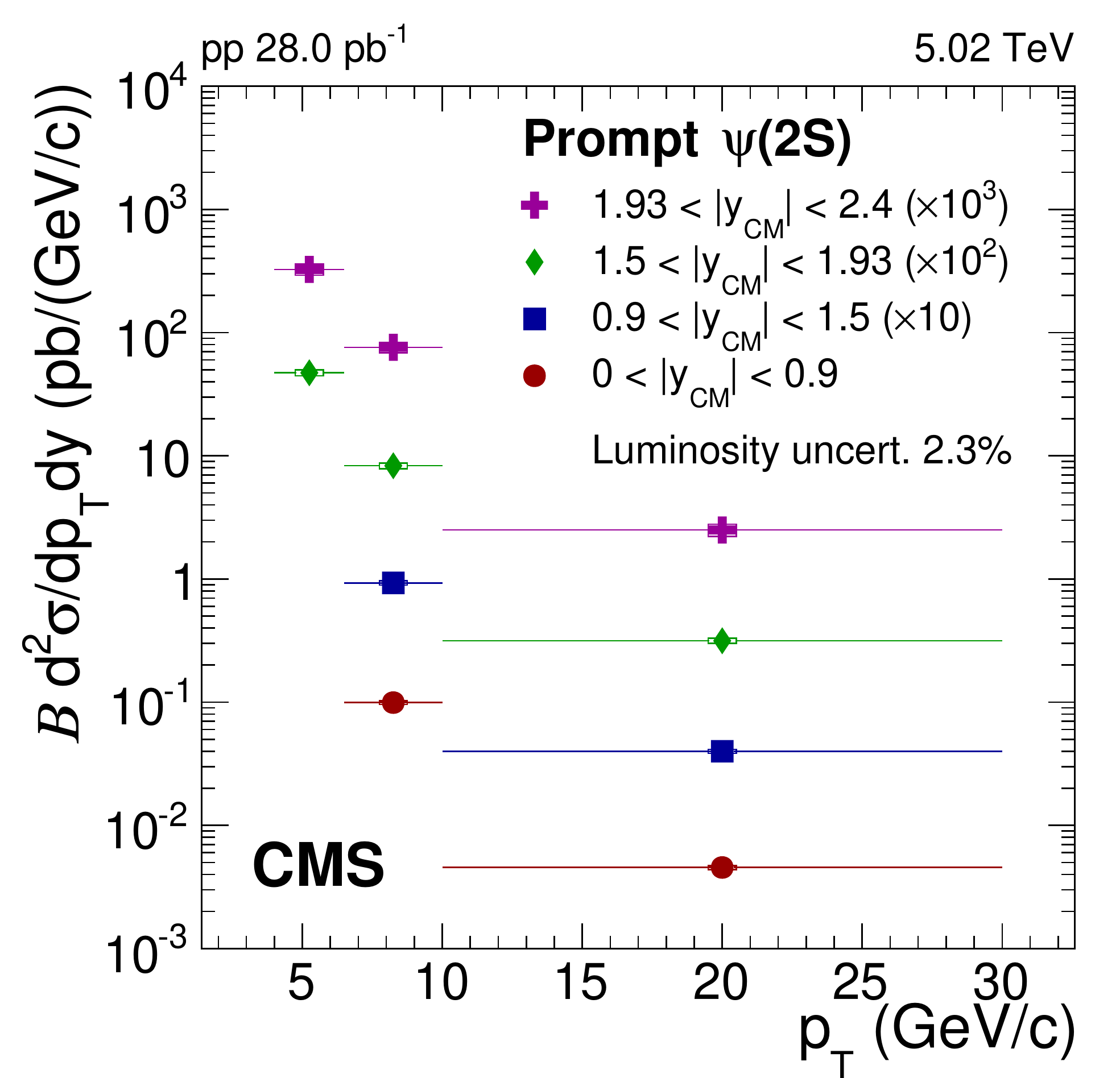}
  \caption{\label{Fig:sigma_pp} Differential cross section (multiplied by the dimuon branching fraction) of prompt \Pgy\ production
    in \pp collisions at {$\sqrts=5.02\TeV$} as a function of \pt,
    in four \ycm bins.
    Statistical and systematic uncertainties are represented with error bars and boxes, respectively. The fully correlated luminosity
    uncertainty of 2.3\% is not included in the point-by-point uncertainties.}
\end{figure}

Figures~\ref{Fig:sigma_ppb} and \ref{Fig:sigma_pp} show the production cross sections of prompt \Pgy\ mesons
in \pPb and \pp collisions at {$\sqrtsNN=5.02\TeV$}, respectively. The results are given as a function of \pt and
in four rapidity bins, separately for forward (the direction of the proton beam) and backward (the direction of the Pb beam) rapidities
in the case of the \pPb measurements.

The second observable considered is the nuclear modification factor, defined as
\begin{linenomath}
  \begin{equation}
    \label{eqn:rpa}
\rpa(\pt,\ycm) \equiv \frac{({\rd^2\sigma}/{\rd\pt\,\rd\ycm})_{\pPb}}{\mathrm{A}({\rd^2\sigma}/{\rd\pt\,\rd\ycm})_{\pp}}.
  \end{equation}
\end{linenomath}
In the absence of any nuclear effects the measured yield in \pPb collisions would be just a superposition of $A=208$ (the number
of nucleons in the Pb nucleus) \pp collisions, and the \rpa would be one.

\begin{figure*}[ht!]
\centering
\includegraphics[width=0.48\textwidth]{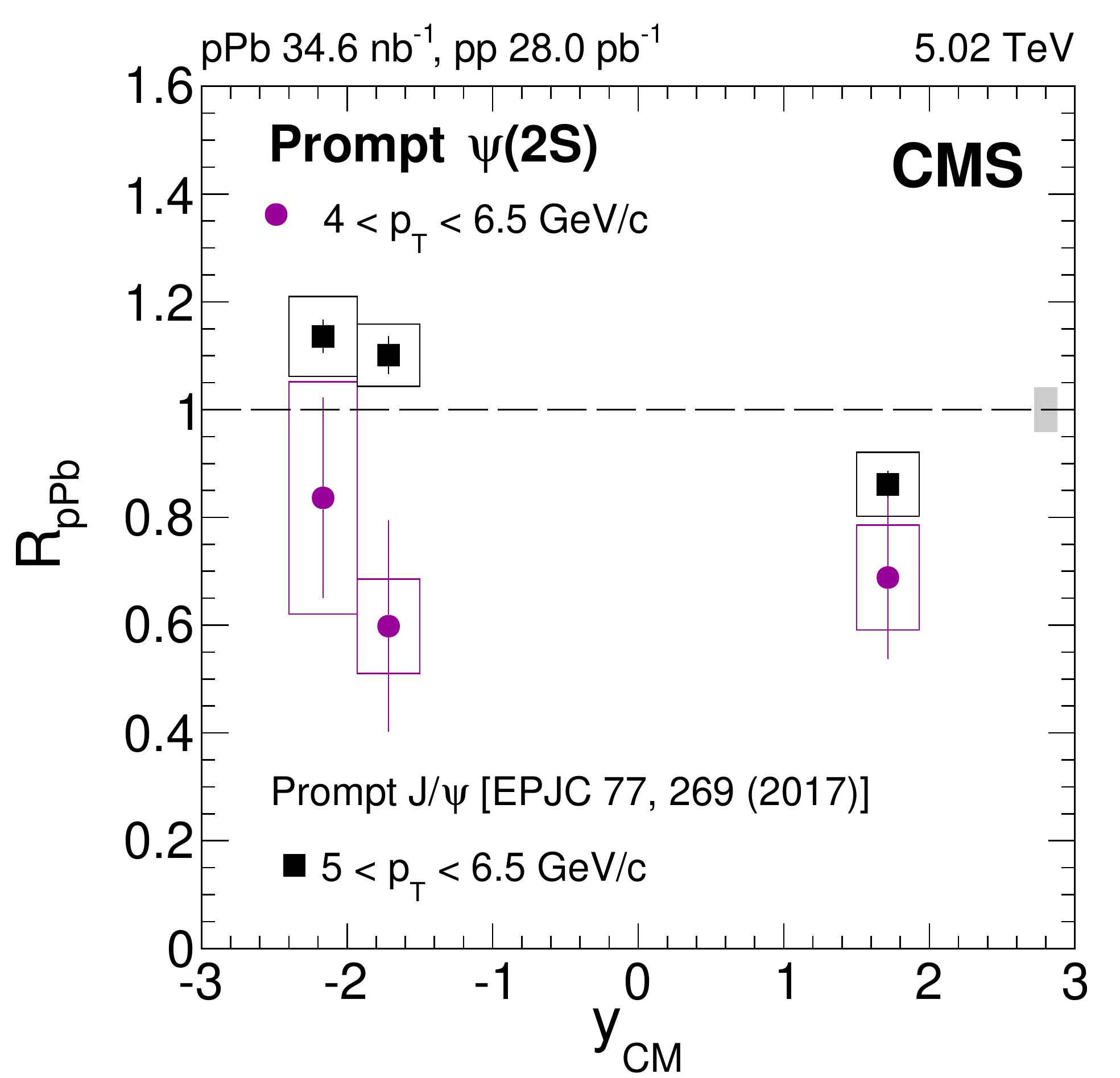}
\includegraphics[width=0.48\textwidth]{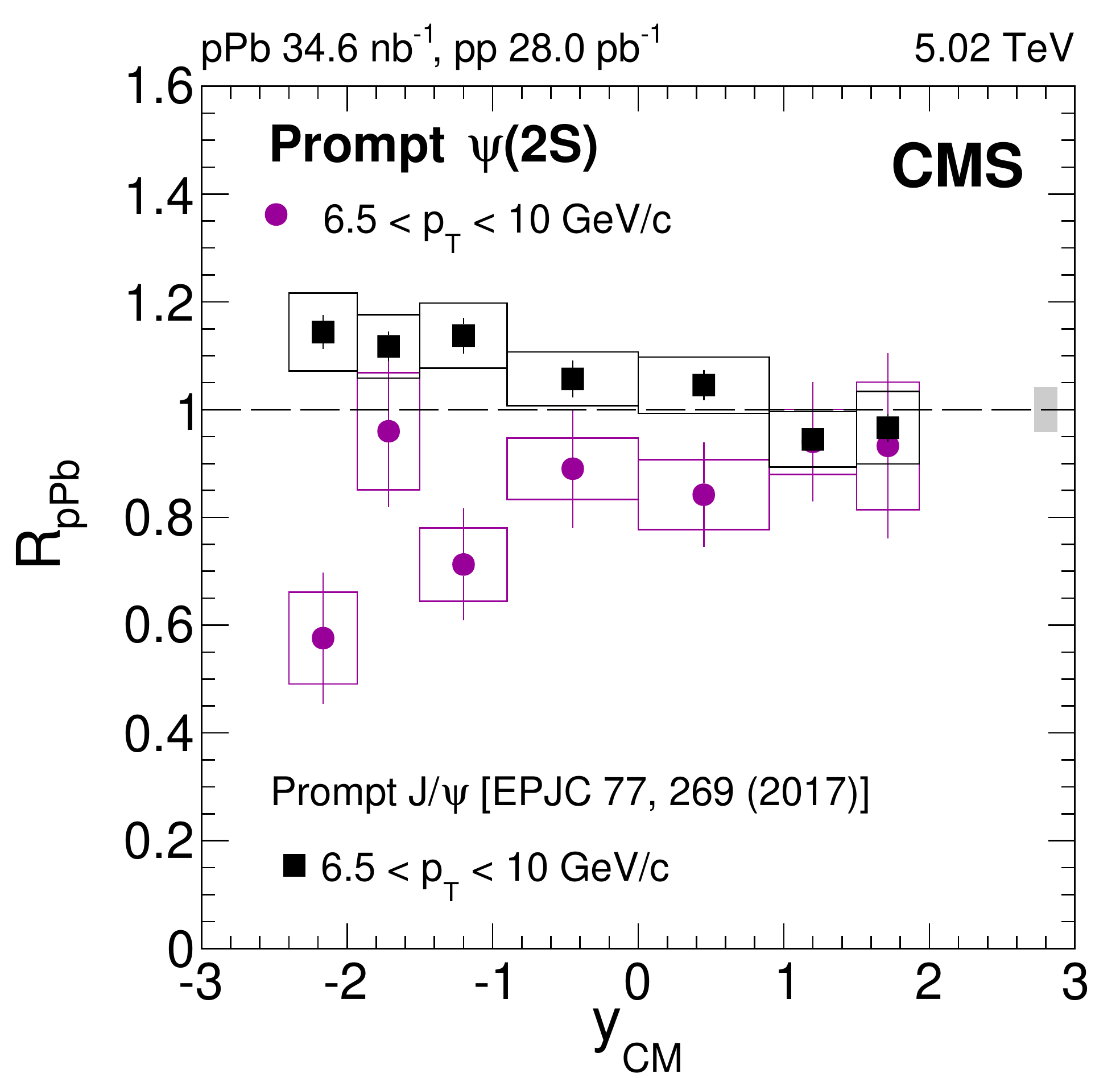}
\includegraphics[width=0.48\textwidth]{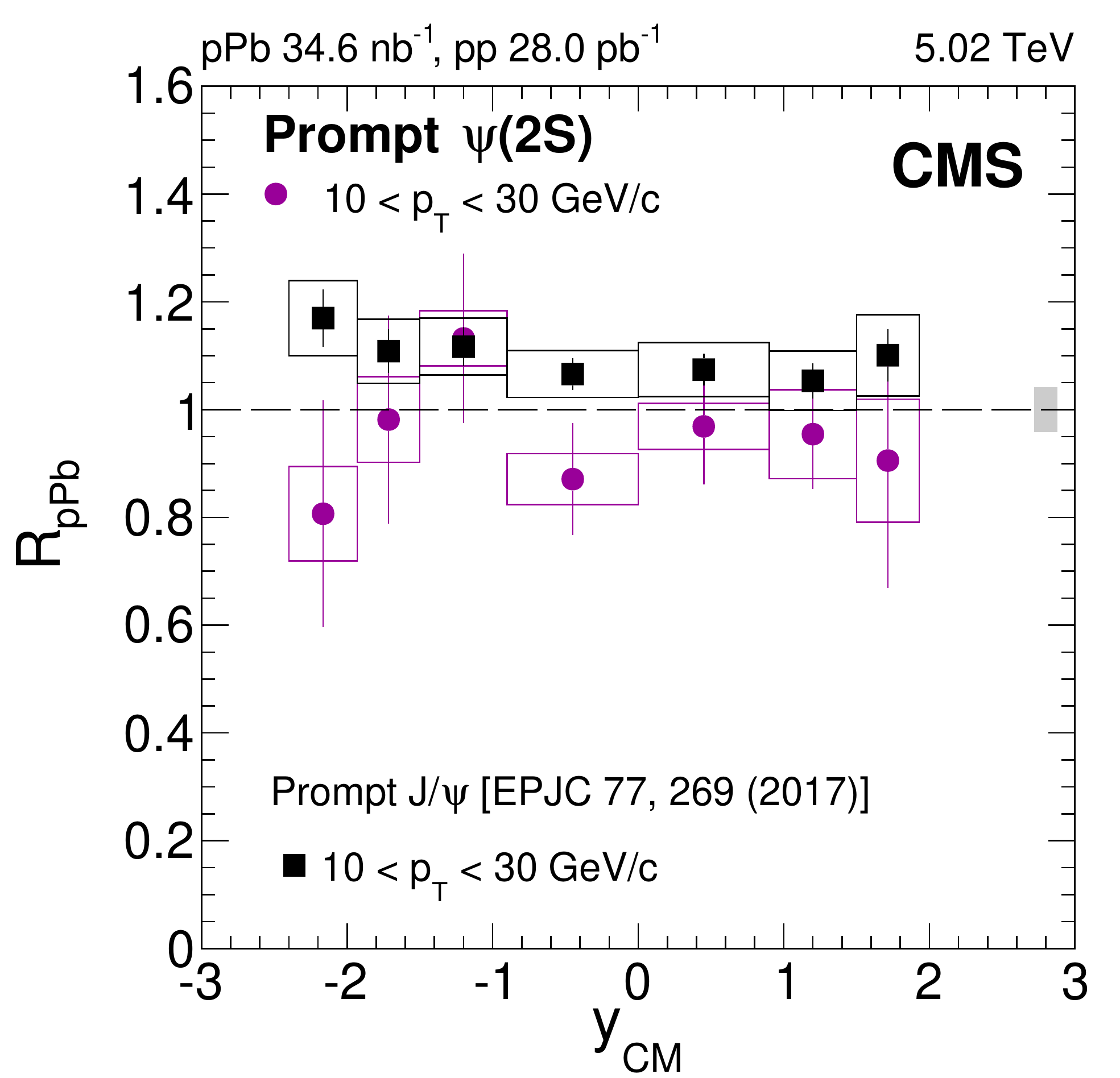}
\caption{\label{Fig:rppb_rap} Rapidity dependence of the
  prompt \Pgy\ \rpa in three \pt ranges. For comparison, the prompt \PJGy\
 nuclear modification factor~\cite{Sirunyan:2017mzd} is also shown.
  Statistical and systematic uncertainties are represented with error bars and boxes, respectively. The fully correlated global
  uncertainty of 4.2\% (that affects both charmonia) is displayed as a box around $\rpa=1$.}
\end{figure*}

Figure~\ref{Fig:rppb_rap} shows the rapidity dependence of the prompt \Pgy\ \rpa in three \pt ranges: {4--6.5}, {6.5--10}, and {10--30}\GeVc.
In the backward rapidity region, the prompt \Pgy\ mesons are suppressed in the intermediate \pt region, and compatible with unity in the rest. In the forward rapidity region, the \rpa is consistent with unity for all \pt bins (although systematically smaller). For comparison, the prompt \PJGy\ nuclear modification factor~\cite{Sirunyan:2017mzd} is also shown in Fig.~\ref{Fig:rppb_rap}. The measured value of \rpa for prompt \Pgy\ mesons, when integrated over \pt and rapidity ({$6.5 < \pt < 30$\GeVc}, {$\abs{y} < 1.6$}), is $0.852\pm 0.037\stat\pm 0.062\syst$. The prompt \PJGy\ \rpa in the same kinematic range is $1.108\pm 0.021\stat\pm 0.055\syst$. The \rpa for prompt \PJGy\ mesons lies systematically above that of the \Pgy\ state. The difference between the two sets of results was quantified, separately for the backward and forward regions, by comparing the two distributions with a $\chi^2$-test, removing the correlated uncertainties from the integrated luminosity. The $p$-values are $<0.05$ for the lowest two \pt ranges in the backward rapidity region, and $>0.37$ in all the other rapidity and \pt intervals. This shows that, with a $p$-value cutoff of 0.05, the two sets of results are incompatible with each other in the backward rapidity region for $\pt<10\GeVc$, but are compatible in the rest. This suggests the presence of different nuclear effects at play in the production of the excited charmonium state compared to the ground state, in the region of backward rapidity and for $\pt<10\GeVc$.

\begin{figure*}[htp]
\centering
\includegraphics[width=0.48\textwidth]{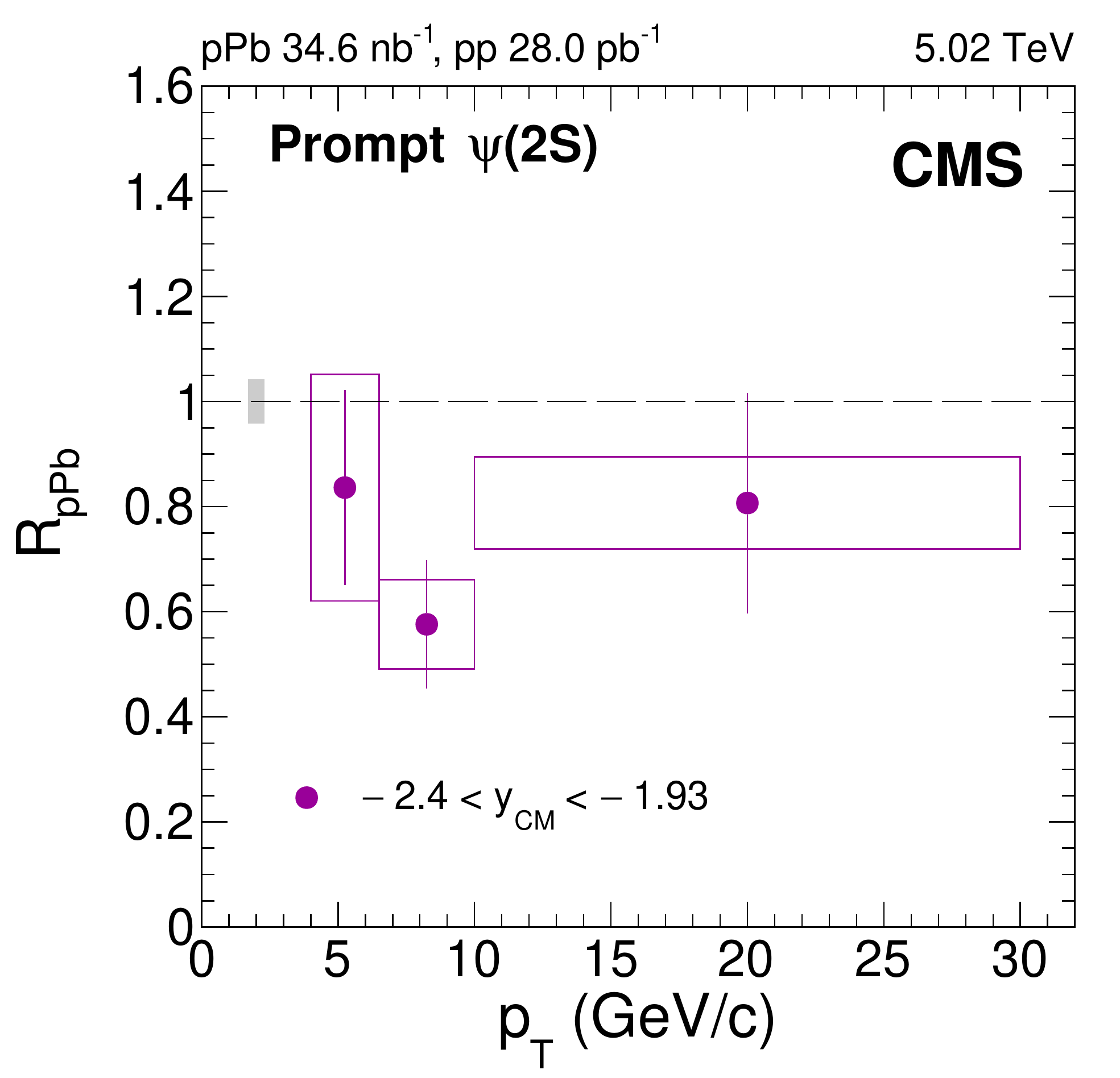}
\includegraphics[width=0.48\textwidth]{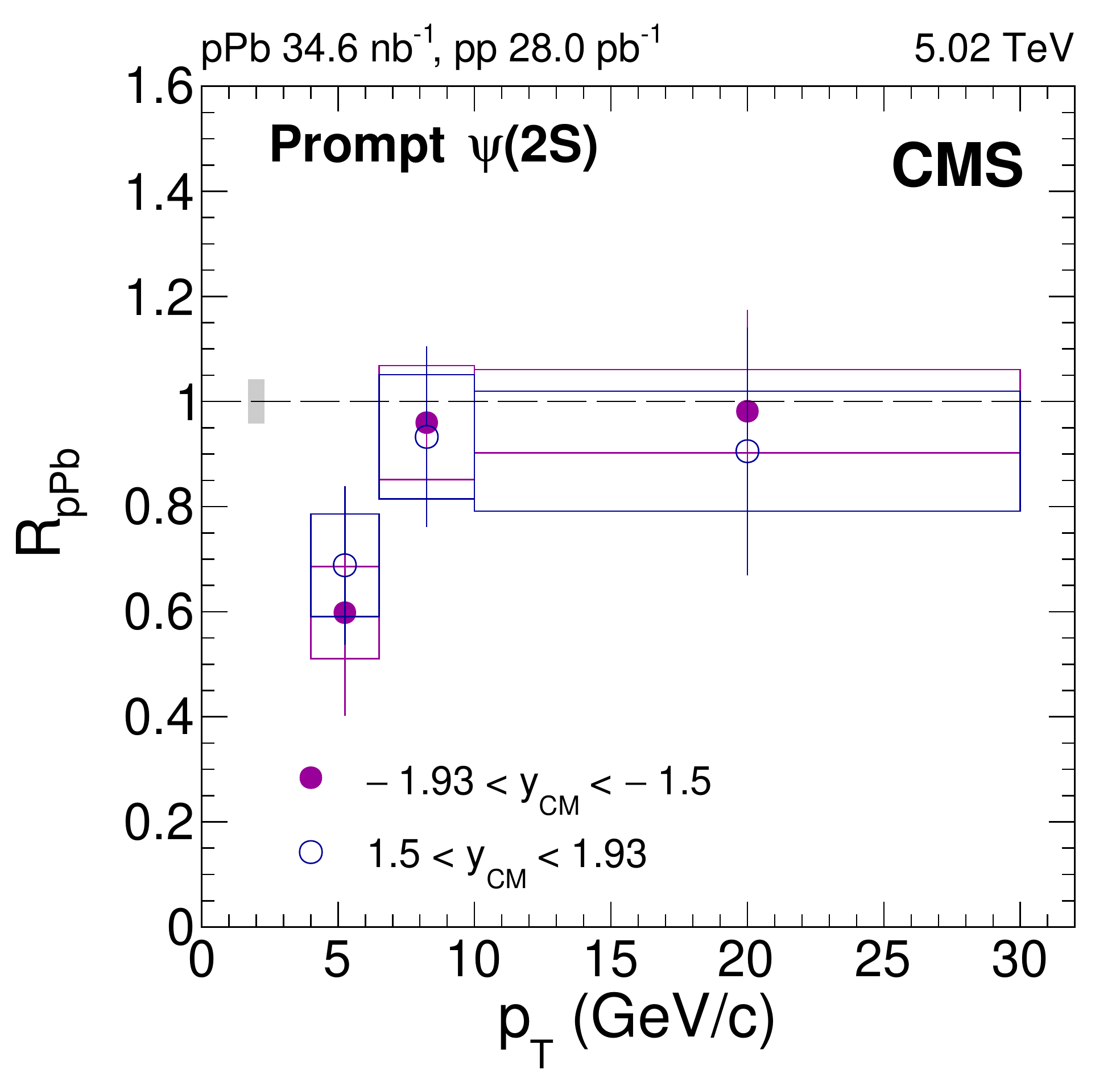}
\includegraphics[width=0.48\textwidth]{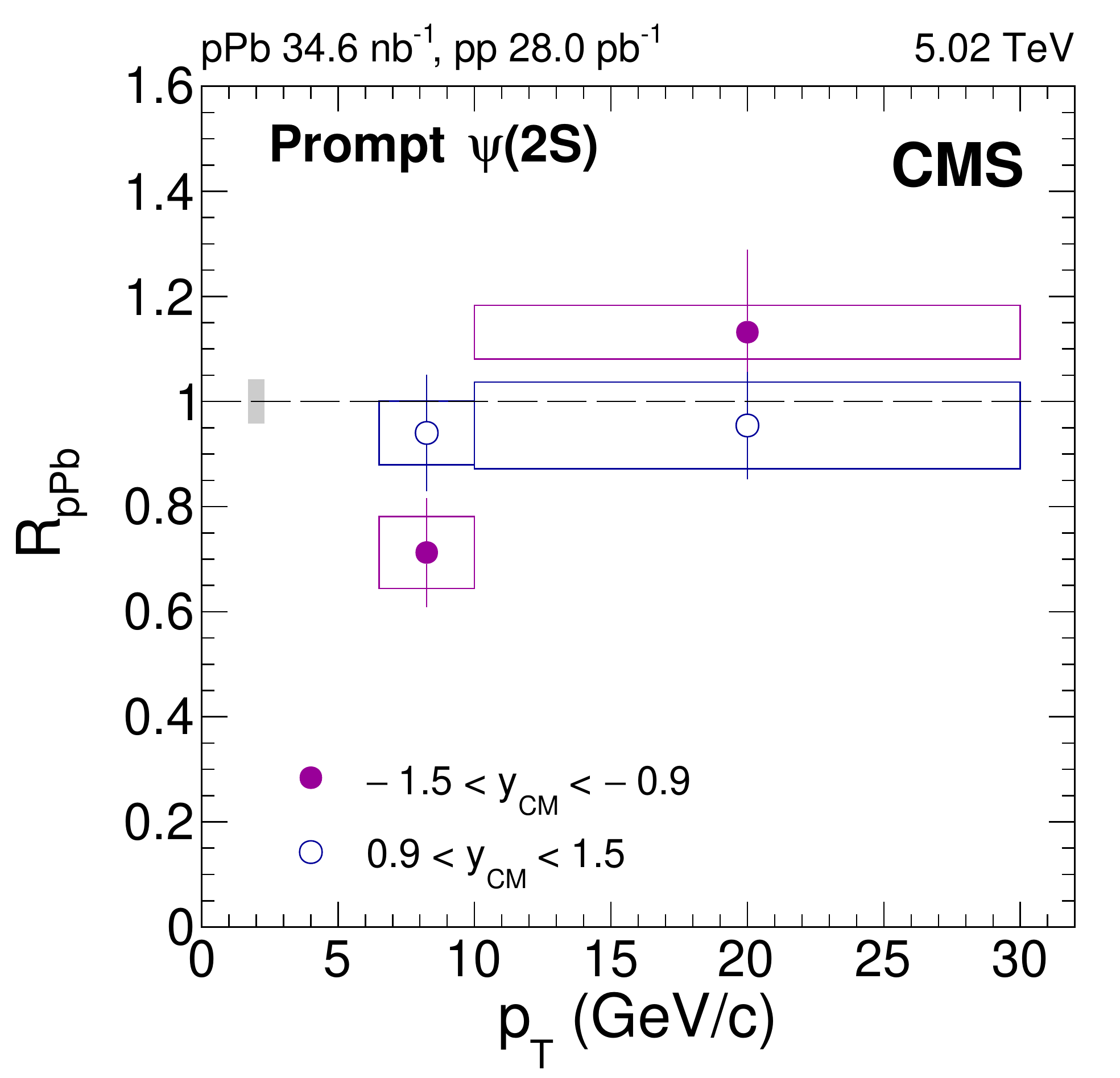}
\includegraphics[width=0.48\textwidth]{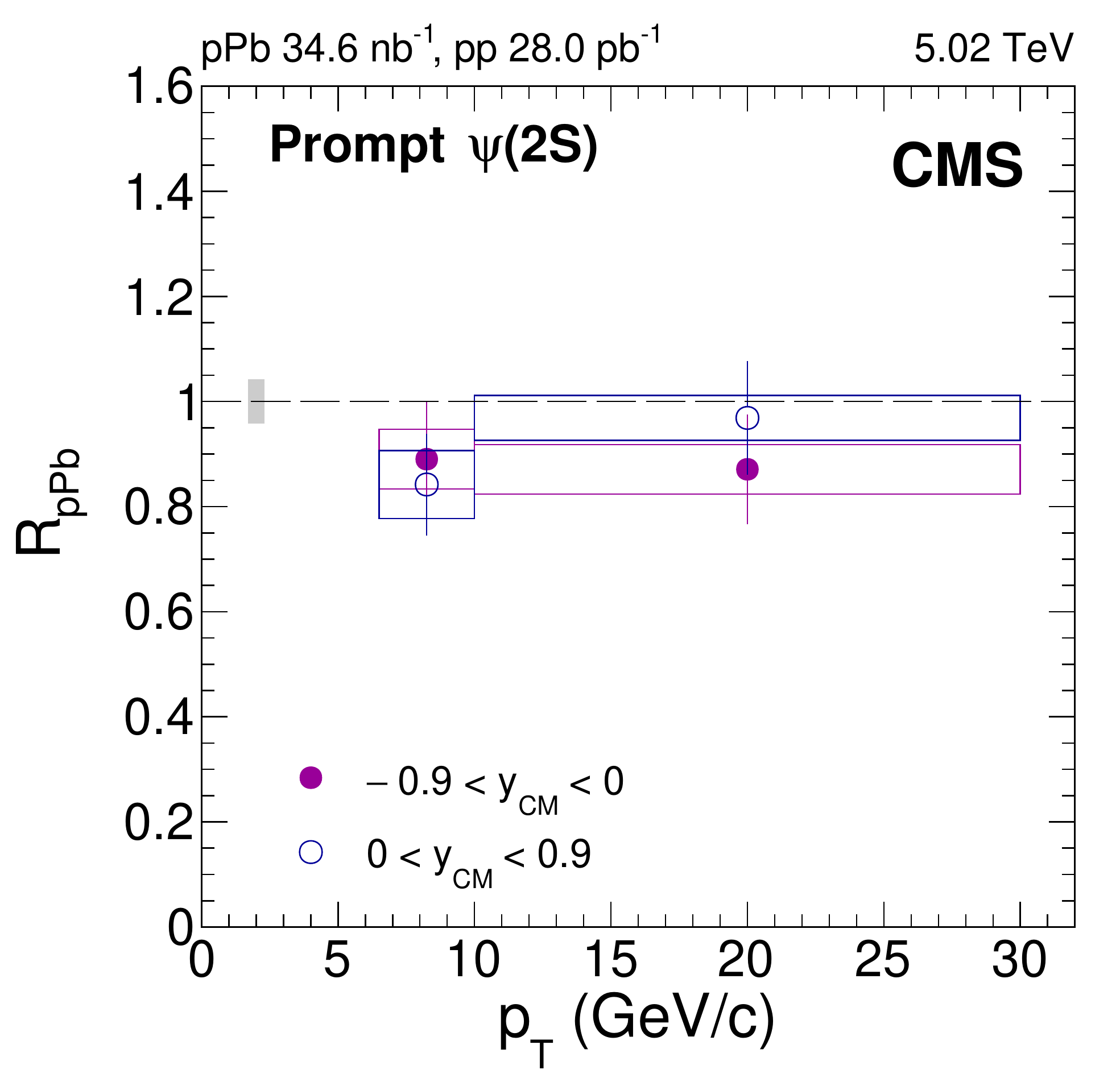}
\caption{\label{Fig:rppb_rap_pt} Transverse momentum dependence of the
  prompt \Pgy\ \rpa in four rapidity ranges.
  Statistical and systematic uncertainties are represented with error bars and boxes, respectively.
   The fully correlated global uncertainty of 4.2\% is displayed as a box
  around {$\rpa=1$}.
}
\end{figure*}

Figure~\ref{Fig:rppb_rap_pt} shows the \pt dependence of the prompt \Pgy\ \rpa in seven rapidity bins, from the most backward $-2.4<\ycm<-1.93$ to the most forward $1.5<\ycm<1.93$ accessible regions. Also in this case, a $\chi^2$-test was used to quantify the significance of the difference of the results from unity, in which the fully correlated global uncertainty, displayed as a box around $\rpa=1$, was not included. The $p$-values are below 0.05 only in the most backward region. This points to the presence of nuclear effects affecting the \Pgy\  production in the most backward region that persists at large transverse momentum.

The suppression of prompt \Pgy\ mesons as compared to prompt \PJGy\
mesons, seen in Fig.~\ref{Fig:rppb_rap}, is compatible with what was measured by the ALICE~\cite{Abelev:2014zpa}, ATLAS~\cite{Aaboud:2017cif}, and LHCb~\cite{Aaij:2016eyl} Collaborations for \pt integrated charmonium production in \pPb collisions at the same collision energy. While nPDFs~\cite{Ferreiro:2013pua,Vogt:2015uba} and coherent energy loss~\cite{Arleo:2012hn,Arleo:2013zua}  are the most
discussed effects to explain prompt \PJGy\ meson suppression, these two phenomena are expected to affect the \rpa of prompt \PJGy\ and \Pgy\ mesons by a similar amount. On the contrary, the final-state interaction with hadrons moving along with the
produced charmonium might lead to a stronger suppression of the \Pgy\ meson, due to its larger size~\cite{Ferreiro:2014bia,Ma:2017rsu}.
The present measurements, covering wide transverse momentum ({$4 < \pt < 30$\GeVc}) and rapidity ($-2.4 < \ycm < 1.93$) ranges, will help understand the origin of the suppression of excited quarkonium states in \pPb collisions at the LHC.

\section{Summary}
The data collected by CMS in \pp and \pPb collisions at $\sqrtsNN=5.02\TeV$
are used to investigate prompt \Pgy\ meson production.
The results are based on data samples corresponding to integrated luminosities of
{28.0\pbinv} for \pp collisions and {34.6\nbinv} for \pPb collisions.
The nuclear modification factor (\rpa) of prompt \Pgy, in the
kinematic range {$4 < \pt < 30$\GeVc} and  $-2.4 < \ycm < 1.93$,
is determined and compared to that of prompt \PJGy\ mesons, reported in Ref.~\cite{Sirunyan:2017mzd}. The prompt \Pgy\ production is suppressed in the intermediate 6.5--10\GeVc \pt interval in the region of backward rapidity. The \rpa is consistent with unity everywhere else (although systematically smaller). The \rpa values of prompt \PJGy\, when compared to those of prompt \Pgy\ mesons, point to different nuclear effects at play in the production of the excited charmonium state compared to the ground state, in the region of backward rapidity and for $\pt<10\GeVc$. The effects of nuclear parton distribution functions or coherent energy loss are expected to affect the \rpa of prompt \PJGy\ and \Pgy\ by a similar amount, thus the results hint the presence of final-state interactions with the medium (partonic or hadronic) produced in \pPb collisions.

\begin{acknowledgments}
We congratulate our colleagues in the CERN accelerator departments for the excellent performance
of the LHC and thank the technical and administrative staffs at CERN and at other CMS institutes for
their contributions to the success of the CMS effort. In addition, we gratefully acknowledge the
computing centers and personnel of the Worldwide LHC Computing Grid for delivering so effectively the
computing infrastructure essential to our analyses.
Finally, we acknowledge the enduring support for the construction and operation of the LHC and the CMS
detector provided by the following funding agencies: BMWFW and FWF (Austria); FNRS and FWO (Belgium);
CNPq, CAPES, FAPERJ, and FAPESP (Brazil); MES (Bulgaria); CERN; CAS, MoST, and NSFC (China);
COLCIENCIAS (Colombia); MSES and CSF (Croatia); RPF (Cyprus); SENESCYT (Ecuador); MoER, ERC IUT, and ERDF (Estonia); Academy of Finland, MEC, and HIP (Finland); CEA and CNRS/IN2P3 (France); BMBF, DFG, and HGF (Germany); GSRT (Greece); NKFIA (Hungary); DAE and DST (India); IPM (Iran); SFI (Ireland); INFN (Italy); MSIP and NRF (Republic of Korea); LAS (Lithuania); MOE and UM (Malaysia); BUAP, CINVESTAV, CONACYT, LNS, SEP, and UASLP-FAI (Mexico); MBIE (New Zealand); PAEC (Pakistan); MSHE and NSC (Poland); FCT (Portugal); JINR (Dubna); MON, RosAtom, RAS and RFBR (Russia); MESTD (Serbia); SEIDI, CPAN, PCTI and FEDER (Spain); Swiss Funding Agencies (Switzerland); MST (Taipei); ThEPCenter, IPST, STAR,
and NSTDA (Thailand); TUBITAK and TAEK (Turkey); NASU and SFFR (Ukraine); STFC (United Kingdom);
DOE and NSF (USA).

\hyphenation{Rachada-pisek} Individuals have received support from the Marie-Curie program and the European Research Council and Horizon 2020 Grant, contract No. 675440 (European Union); the Leventis Foundation; the A. P. Sloan Foundation; the Alexander von Humboldt Foundation; the Belgian Federal Science Policy Office; the Fonds pour la Formation \`a la Recherche dans l'Industrie et dans l'Agriculture (FRIA-Belgium); the Agentschap voor Innovatie door Wetenschap en Technologie (IWT-Belgium); the F.R.S.-FNRS and FWO (Belgium) under the ``Excellence of Science - EOS" - be.h project n. 30820817; the Ministry of Education, Youth and Sports (MEYS) of the Czech Republic; the Lend\"ulet (``Momentum") Program and the J\'anos Bolyai Research Scholarship of the Hungarian Academy of Sciences, the New National Excellence Program \'UNKP, the NKFIA research grants 123842, 123959, 124845, 124850 and 125105 (Hungary); the Council of Science and Industrial Research, India; the HOMING PLUS program of the Foundation for Polish Science, cofinanced from European Union, Regional Development Fund, the Mobility Plus program of the Ministry of Science and Higher Education, the National Science Center (Poland), contracts Harmonia 2014/14/M/ST2/00428, Opus 2014/13/B/ST2/02543, 2014/15/B/ST2/03998, and 2015/19/B/ST2/02861, Sonata-bis 2012/07/E/ST2/01406; the National Priorities Research Program by Qatar National Research Fund; the Programa Estatal de Fomento de la Investigaci{\'o}n Cient{\'i}fica y T{\'e}cnica de Excelencia Mar\'{\i}a de Maeztu, grant MDM-2015-0509 and the Programa Severo Ochoa del Principado de Asturias; the Thalis and Aristeia programs cofinanced by EU-ESF and the Greek NSRF; the Rachadapisek Sompot Fund for Postdoctoral Fellowship, Chulalongkorn University and the Chulalongkorn Academic into Its 2nd Century Project Advancement Project (Thailand); the Welch Foundation, contract C-1845; and the Weston Havens Foundation (USA).
\end{acknowledgments}

\bibliography{auto_generated}

\cleardoublepage \appendix\section{The CMS Collaboration \label{app:collab}}\begin{sloppypar}\hyphenpenalty=5000\widowpenalty=500\clubpenalty=5000\vskip\cmsinstskip
\textbf{Yerevan Physics Institute, Yerevan, Armenia}\\*[0pt]
A.M.~Sirunyan, A.~Tumasyan
\vskip\cmsinstskip
\textbf{Institut f\"{u}r Hochenergiephysik, Wien, Austria}\\*[0pt]
W.~Adam, F.~Ambrogi, E.~Asilar, T.~Bergauer, J.~Brandstetter, E.~Brondolin, M.~Dragicevic, J.~Er\"{o}, A.~Escalante~Del~Valle, M.~Flechl, M.~Friedl, V.M.~Ghete, J.~Hrubec, M.~Jeitler\cmsAuthorMark{1}, N.~Krammer, I.~Kr\"{a}tschmer, D.~Liko, T.~Madlener, I.~Mikulec, N.~Rad, H.~Rohringer, J.~Schieck\cmsAuthorMark{1}, R.~Sch\"{o}fbeck, M.~Spanring, D.~Spitzbart, A.~Taurok, W.~Waltenberger, J.~Wittmann, C.-E.~Wulz\cmsAuthorMark{1}, M.~Zarucki
% ...
\vskip\cmsinstskip
% ...
1:  Also at Vienna University of Technology, Vienna, Austria\\
% ...
\end{sloppypar}
\end{document}